\documentclass[12pt]{article}

\usepackage{epsfig}

\def\NPB{{\em Nucl. Phys.} B~}
\def\PLB{{\em Phys. Lett.} B~}
\def\PRL{{\em Phys. Rev. Lett.}~}
\def\PRC{{\em Phys. Rev.} C~}
\def\PRD{{\em Phys. Rev.} D~}

\newcommand{\be}{\begin{eqnarray}}
\newcommand{\ee}{\end{eqnarray}}

\def\lsim{\mathrel{\rlap{\lower4pt\hbox{\hskip1pt$\sim$}}
    \raise1pt\hbox{$<$}}}               
\def\gsim{\mathrel{\rlap{\lower4pt\hbox{\hskip1pt$\sim$}}
    \raise1pt\hbox{$>$}}}               

\begin{document}

\rightline{{\Large Preprint RM3-TH/02-17}}

\vspace{2cm}

\begin{center}

\LARGE{Proton and neutron polarized structure functions\\[2mm] from low to high $Q^2$ \footnote{To appear in the Proceedings of the II International Symposium on the {\em Gerasimov-Drell-Hearn sum rule and the spin structure of the nucleon}, Genova (Italy), July 3-6, 2002, World Scientific Publishing (Singapore), in press.}}

\vspace{1cm}

\large{S. Simula$^{(*)}$, M. Osipenko$^{(**)}$, G. Ricco$^{(***)}$ and M. Taiuti$^{(***)}$}\\

\vspace{0.5cm}

\normalsize{$^{(*)}$Istituto Nazionale di Fisica Nucleare, Sezione Roma III,\\ Via della Vasca Navale 84, I-00146 Roma, Italy\\$^{(**)}$Physics Department, Moscow State University, 119899 Moscow, Russia\\$^{(***)}$Dipartimento di Fisica, Universit\'a di Genova and INFN, Sezione di Genova,\\ Via Dodecanneso 33, I-16146, Genova, Italy}

\end{center}

\vspace{1cm}

\begin{abstract}

\noindent Phenomenological parameterizations of proton and neutron polarized structure functions, $g_1^p$ and $g_1^n$, are developed for $x \gsim 0.02$ using deep inelastic data up to $\sim 50 ~ (GeV/c)^2$ as well as available experimental results on photo- and electro-production of nucleon resonances. The generalized Drell-Hearn-Gerasimov sum rules are predicted from low to high values of $Q^2$ and compared with proton and neutron data. Furthermore, the main results of the power correction analysis carried out on the $Q^2$-behavior of the polarized proton Nachtmann moments, evaluated using our parameterization of $g_1^p$, are briefly summarized.

\end{abstract}

\newpage

\pagestyle{plain}

\section{Introduction}

\indent The experimental investigation of lepton deep-inelastic scattering ($DIS$) off proton and deuteron targets has provided a wealth of information on parton distributions in the nucleon. In the past few years some selected issues in the kinematical regions corresponding to large values of the Bjorken variable $x$ have attracted a lot of theoretical and phenomenological interest; among them one should mention the occurrence of power corrections associated to {\em dynamical} higher-twist operators measuring the correlations among partons. The extraction of the latter is of particular relevance since the comparison with theoretical predictions either based on lattice $QCD$ simulations or obtained from models of the nucleon structure represents an important test of $QCD$ in its non-perturbative regime.

\indent In Refs.~\cite{Ricco} and \cite{SIM00} phenomenological fits of the world data on the unpolarized nucleon structure functions $F_2^N$ and $F_L^N$ were used to evaluate Nachtmann moments and power correction analyses were carried out. In Ref.~\cite{SIM02} the same twist analysis was extended to the polarized proton case by developing a parameterization of $g_1^p$, which describes the $DIS$ proton data up to $Q^2 \sim 50 ~ (GeV/c)^2$ and includes a phenomenological Breit-Wigner ans\"atz able to reproduce the existing electroproduction data in the proton-resonance regions. The interpolation formula for $g_1^p$ was successfully extended down to the photon point, showing that it nicely reproduces the Mainz data~\cite{Mainz} on the energy dependence of the asymmetry of the transverse photoproduction cross section.

\indent The plan of this contribution is as follows. In Section 2 we extend our phenomenological parameterization of Ref.~\cite{SIM02} to the neutron polarized structure function $g_1^n$. In Section 3 the generalized Drell-Hearn-Gerasimov ($DHG$) sum rules are predicted from low to high values of $Q^2$ and compared with proton and neutron data. Finally, in Section 4 the main results of the power correction analysis carried out in Ref.~\cite{SIM02} on the $Q^2$-behavior of the polarized proton Nachtmann moments will be briefly summarized.

\section{Phenomenological parameterizations of $g_1^p$ and $g_1^n$ from low to high $Q^2$}

\indent Following Ref.~\cite{SIM02} we write the polarized nucleon structure functions as the sum of three contributions 
  \be
     g_i(x, Q^2) = g_i^{(el.)}(x, Q^2) + g_i^{(res.)}(x, Q^2) +  
     g_i^{(non-res.)}(x, Q^2) 
    \label{eq:gi}
 \ee
where the suffix $p$ or $n$ is omitted for simplicity, $g_i^{(el.)}$, $g_i^{(res.)}$ and $g_i^{(non-res.)}$ are the elastic, resonant and non-resonant contributions to $g_i$, respectively, and $i = 1, 2$. In Eq. (\ref{eq:gi}) possible interference terms between the resonant and non-resonant contributions are neglected, since they are well beyond the scope of our phenomenological fit.

\indent The elastic contribution is the simplest one, because it can be expressed in terms of the nucleon Sachs form factors (see Eqs.~(17-18) of Ref.~\cite{SIM02}). As for the non-resonant terms $g_i^{(non-res.)}$ we adopt the following decomposition
\be
    g_1^{(non-res.)}(x, Q^2) & = & g^{\Delta \sigma}(x, Q^2) + {4 M^2 x^2 
    \over Q^2} g^{LT}(x, Q^2)  \nonumber \\
    g_2^{(non-res.)}(x, Q^2) & = & - g^{\Delta \sigma}(x, Q^2) + 
    g^{LT}(x, Q^2)
    \label{eq:g12_NR}
 \ee
where $g^{\Delta \sigma}$ is the contribution arising from the transverse asymmetry $A_1$, while $g^{LT}$ is the $LT$ interference governing the asymmetry $A_2$. In Ref.~\cite{SIM02} we have developed an interpolation formula for $g^{\Delta \sigma}$ of the form
 \be
     g^{\Delta \sigma} = {W^2 - M^2 \over 2W^2} \sum_{j = 1}^N a_j \left[ 1 
     + {W^2 \over Q^2 + Q_R^2} \right]^{\alpha_j} \left[ {W^2 - W_{\pi}^2 
     \over W^2 - W_{\pi}^2 + Q^2 + W_T^2} \right]^{\beta_j} ~~~~
    \label{eq:g_Delta}
 \ee
where $W_{\pi}$ is the pion production threshold and $t$ is a parameter aimed at describing the logarithmic scaling violations in the $DIS$ regime, namely: $t = \mbox{ln} \left\{ \mbox{ln} \left[ (Q^2 + Q_0^2) / \Lambda^2 \right] / \mbox{ln} \left( Q_0^2 / \Lambda^2 \right) \right\}$. In Eq.~(\ref{eq:g_Delta}) the parameter $Q_R^2$ describes the transition from the expected dominance of the Regge behavior at $Q^2 \lsim Q_R^2$ to the partonic regime at $Q^2 >> Q_R^2$, while the quantities $a_j$, $\alpha_j$ and $\beta_j$ are parameters assumed to depend linearly on $t$. Finally, the term $g^{LT}$, contributing to Eq. (\ref{eq:g12_NR}), is parameterized as in Eq.~(26) of Ref.~\cite{SIM02}.

\indent All the parameters appearing in Eq.~(\ref{eq:g_Delta}) but $W_T$ can be determined by fitting existing measurements of the asymmetry $A_1$ in the $DIS$ kinematics ($W \gsim 2 ~ GeV$). As explained in Ref.~\cite{SIM02} the value of $W_T$ can be fixed by requiring the reproduction of the $DHG$ sum rule. We point out that in Eq. (\ref{eq:g_Delta}) $g^{\Delta \sigma}$ is assumed to behave in the Bjorken limit as a power of $x$ at low $x$. There is no strong argument in favor of such an assumption and therefore Eq. (\ref{eq:g_Delta}) has to be considered as a simple approximation valid in a limited $x$-range. In this respect, since existing data for both proton and neutron targets are scarce below $x \sim 0.02$, we consider $x \gsim 0.02$ as the $x$-range of applicability of our parameterization (\ref{eq:g_Delta}). This implies that we cannot check the Bjorken sum rule, because the latter is extremely sensitive to the behavior of $g_1^n$ at very low $x$ (below $10^{-2}$). Therefore, in this contribution the parameterization of $g_1^p$ is directly taken from Ref.~\cite{SIM02}, and the theoretical value of the Bjorken sum rule is inserted in the fitting data set for $g_1^n$.

\indent In case of proton we used~\cite{SIM02} $14$ parameters against $209$ experimental points, obtaining for the $\chi^2$ variable (divided by the number of {\em d.o.f.}) the minimum value of $0.66$. Repeating the same procedure in case of the neutron (with the addition of the Bjorken sum rule as a constraint) we have got $\chi^2 = 0.90$ against the $245$ experimental points from Refs.~\cite{data_D,data_n}. We anticipate that the value of the parameter $W_T$, fixed through the $DHG$ sum rule, turns out to be $W_T = 0.475 ~ (0.451) ~ GeV$ in case of proton (neutron).

\begin{figure}[t]

\centerline{\epsfxsize=16cm \epsfbox{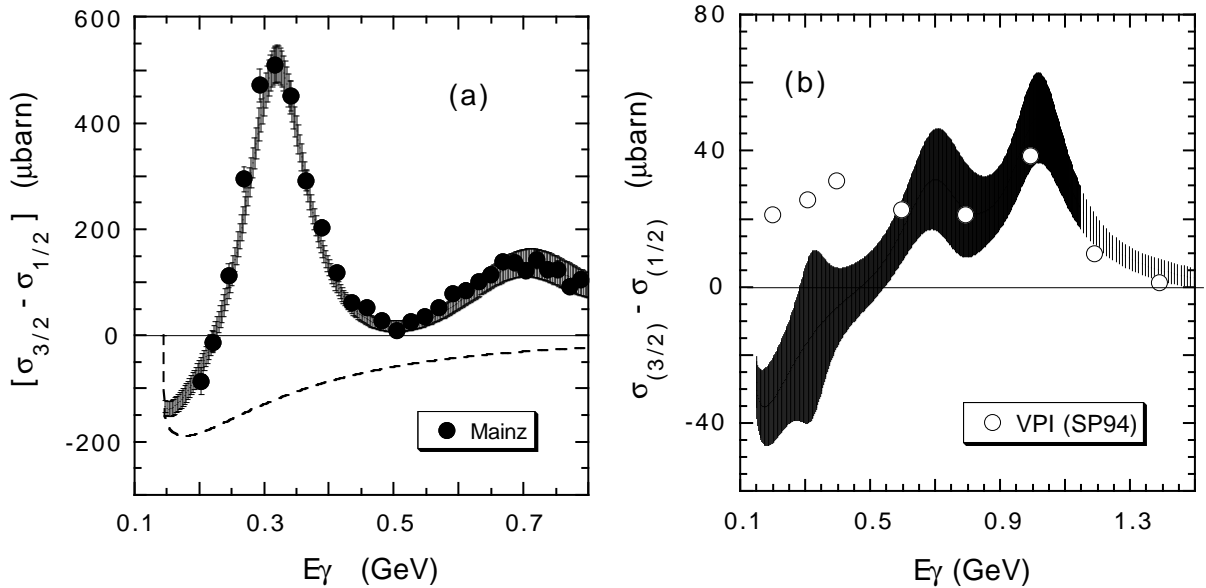}}

\caption{Asymmetry of the transverse photoabsorption cross section for the proton (a) and for the difference between proton and neutron (b) versus the photon energy $E_{\gamma}$. Full dots are the data from Ref.~\protect\cite{Mainz}, while open dots are the results of the $VPI$ multipole analysis labelled $SP94$ (see Ref.~\protect\cite{VPI}). The shaded area is our prediction, while the dashed line in (a) is our non-resonant contribution, which is sensitive to the value of the parameter $W_T$ (see text). Fig.~1(a) is adapted from Ref.~\protect\cite{SIM02}.}

\end{figure}

\indent In the resonance regions ($W \lsim 2 ~ GeV$) we adopt a simple Breit-Wigner shape to describe the $W$-dependence of the contribution of an isolated resonance $R$, while its $Q^2$-dependence can be conveniently expressed in terms of the helicity amplitudes $A_{1/2}^R$, $A_{3/2}^R$ and $S_{1/2}^R$. The explicit expression for $g_1^{(res.)}$ and $g_2^{(res.)}$ are given by Eqs.~(29-35) of Ref.~\cite{SIM02}. We have considered all the "four-star" resonances of the $PDG$ \cite{PDG} having a mass $M_R < 2 ~ GeV$ and a total transverse photoamplitude $\sqrt{|A_{1/2}^R|^2 + |A_{3/2}^R|^2}$ larger than $0.050 ~ (0.040) ~ GeV^{-1/2}$ in case of proton (neutron). Our final results at $Q^2 = 0$ are reported in Fig.~1. It can be seen that they positively compare with all the Mainz data \cite{Mainz} in case of the proton and with the results of the $VPI$ multipole analysis \cite{VPI} for photon energies $E_{\gamma}$ above $\approx 0.5 ~ GeV$ in case of the isovector-isoscalar ($VS$) channel. There is however a clear discrepancy when $E_{\gamma} \lsim 0.5 ~ GeV$. It should be reminded that the results of multipole analyses are consistent with the $VV$ part of the $DHG$ sum rule, while they differ remarkably in case of the $VS$ part.

\section{Generalized $DHG$ sum rules from low to high $Q^2$}

\indent In this Section we present our predictions for the generalized $DHG$ sum rules based on the parameterizations of $g_1^p$ and $g_1^n$ described previously. 

\indent We have calculated the inelastic part of the first moment of $g_1$, defined as $\Gamma_1(Q^2) \equiv \int_0^{x_{\pi}} dx ~ g_1(x, Q^2)$, where $x_{\pi}$ is the pion threshold. Our results are shown in Fig.~2 and compared with both $DIS$ data and the new $JLab$ data~\cite{Burkert} which cover the intermediate $Q^2$-region ranging from $\approx 0.2$ to $\approx 1 ~ (GeV/c)^2$. It can be seen that our results nicely fit the data in the $DIS$ kinematics and agree at very low values of $Q^2$ with the results of Heavy Baryon Chiral Perturbation Theory ($HB \chi pT$) obtained in Ref.~\cite{Ji}. However, while the proton $JLab$ data change sign at $Q^2 \simeq 0.25 ~ (GeV/c)^2$, our parameterization of $g_1^p$ predicts the occurrence of the zero-crossing point at $Q^2 = 0.16 \pm 0.04 ~ (GeV/c)^2$. Moreover the applicability of the $HB \chi pT$ to the nucleon has been criticized in Ref.~\cite{BHM}, where important finite-mass effects have been found.

\begin{figure}[t]

\centerline{\epsfxsize=16cm \epsfbox{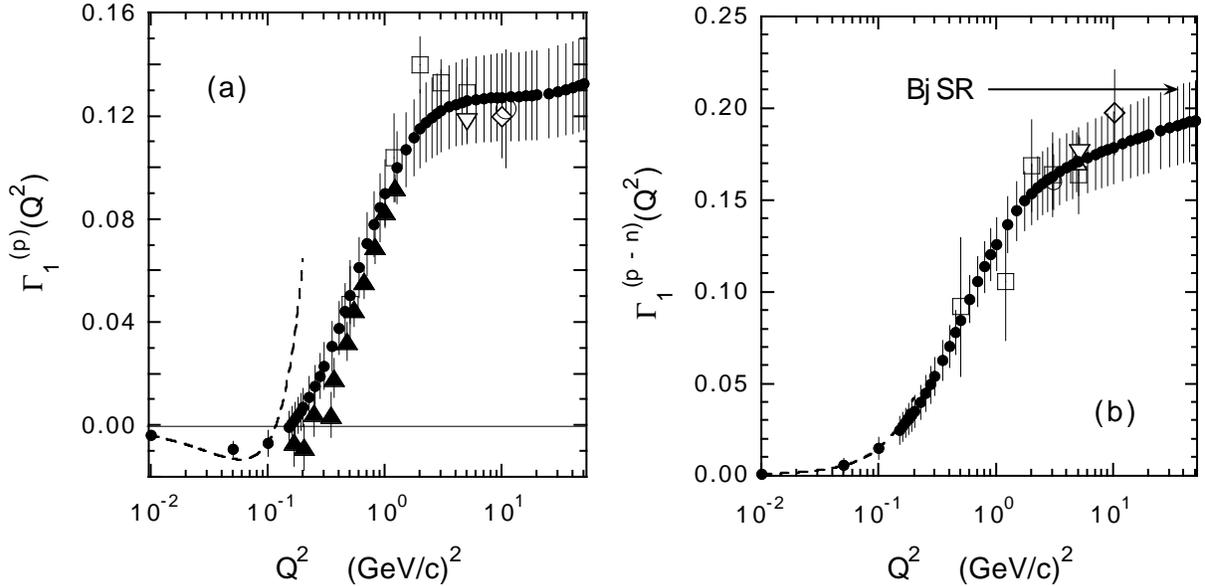}}

\caption{Inelastic part of the first moment of $g_1^p$ (a) and of [$g_1^p - g_1^n$] (b). Full dots represent our results. In (a) full triangles are data from $JLab$~\protect\cite{Burkert}, while open squares, diamonds, dots and triangles are from Ref.~\protect\cite{DIS_p}(a,b,c,d), respectively. In (b) open diamonds, squares and reverse triangles are from Ref.~\protect\cite{data_D}(a,b,c), while open dots and triangles are from Ref.~\protect\cite{data_n}(a,b), respectively. Dashed lines are the $HB \chi pT$ predictions of Ref.~\protect\cite{Ji}.}

\end{figure}

\indent The resonance contribution to the transverse cross section asymmetry of the neutron, defined as $I_t^{(n)}(Q^2) \equiv \int_{\nu_{\pi}}^{\nu_{max}} d\nu ~ (\sigma_{1/2} - \sigma_{3/2}) / \nu$ with $\nu_{max}$ corresponding to $W_{max} \simeq 2 ~ GeV$, has been recently determined by the $JLab$ experiment $E94010$~\cite{JLAB_n}. Our results are reported in Fig.~3 as full dots and compared with the $JLab$ data. It can be seen that a striking discrepancy occurs below $Q^2 \simeq 0.4 ~ (GeV/c)^2$. A similar discrepancy is shared also by the result of the unitary isobar model of Ref.~\cite{DKT}.

\begin{figure}[htb]

\centerline{\epsfxsize=16cm \epsfbox{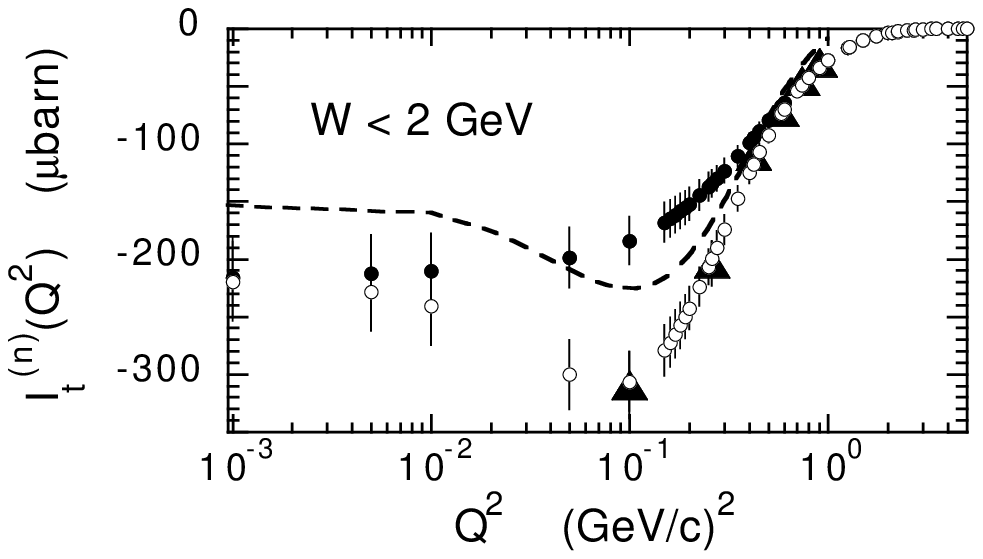}}

\caption{Resonance contribution to the transverse cross section asymmetry of the neutron. Full triangles are data from $JLab$~\protect\cite{JLAB_n}. Open and full dots are our results obtained with and without the low-$Q^2$ modification of the $N-\Delta(1232)$ transition asymmetry (see text). The dashed line is the result of the unitary isobar model of Ref.~\protect\cite{DKT}.}

\end{figure}

\indent Since the $N-\Delta(1232)$ transition dominates at low $Q^2$, the agreement with the neutron $JLab$ data can be recovered by modifying the low-$Q^2$ behavior of the asymmetry $A_1^{\Delta}$. In Fig.~3 the open dots, which nicely fits all the $JLab$ data, correspond to the results of a modified parameterization of $g_1^n$ in which we assume $A_1^{\Delta} = - 0.56 - 0.85 ~ (Q^2 / 0.14) e^{- Q^2 / 0.14}$ instead of the constant value $A_1^{\Delta} = - 0.56$ considered in Ref.~\cite{SIM02} and suggested by all multipole analyses (cf. Ref.~\cite{DKT} and references therein). The modification of $A_1^{\Delta}$ has a direct impact also on our parameterization of $g_1^p$ which should be correspondingly modified. Thus we have recalculated the first moment $\Gamma_1(Q^2)$ and our results are reported in Fig.~4. It can be seen that our modified parameterization of $g_1^p$ predicts the same zero-crossing point of the proton $JLab$ data. Moreover, the low-$Q^2$ behaviors of $\Gamma_1^{(p)}(Q^2)$ and $\Gamma_1^{(p-n)}(Q^2)$ agree very well with the $\chi pT$ prediction of Ref.~\cite{BHM}.

\begin{figure}[htb]

\centerline{\epsfxsize=16cm \epsfbox{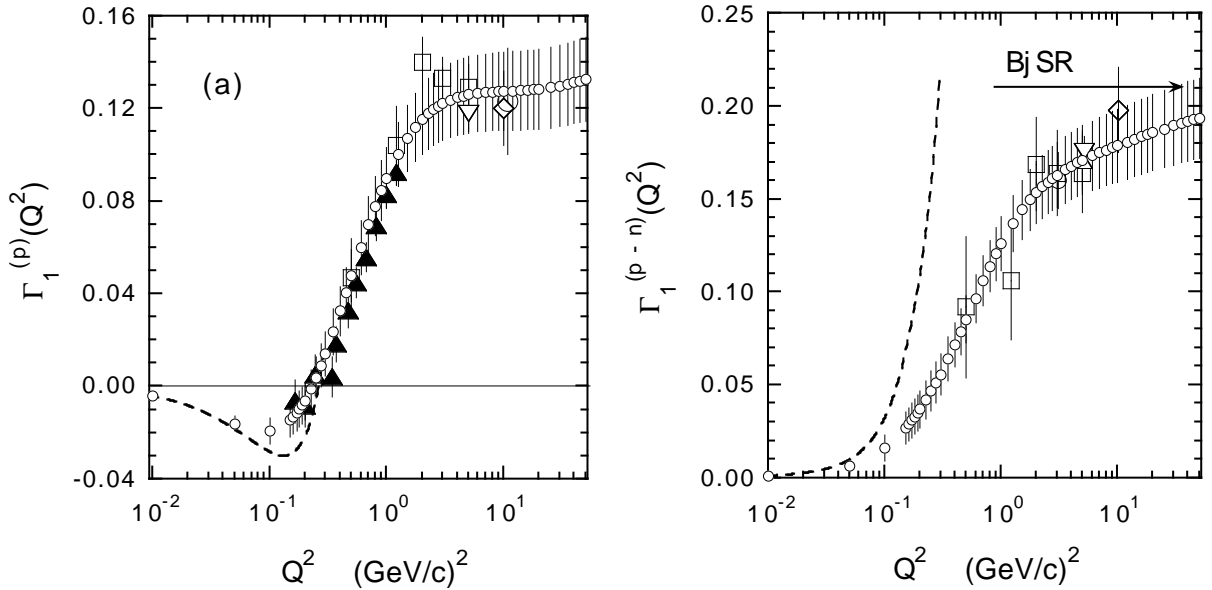}}

\caption{The same as in Fig.~2, with the open dots being our results obtained with the low-$Q^2$ modification of the $N-\Delta(1232)$ transition asymmetry described in the text. The dashed lines are the $\chi pT$ predictions of Ref.~\protect\cite{BHM}.}

\end{figure}

\indent To sum up, the results we have obtained for the generalized $DHG$ sum rules suggest that more work is needed in order to properly parameterize the low-$Q^2$ behavior of the $N-\Delta(1232)$ transition.

\section{Twist analysis of the proton Nachtmann moments}

\indent In this Section we briefly summarize the main results of Ref.~\cite{SIM02} concerning the power correction analysis of the polarized proton Nachtmann moments, $M_n^{(1)}(Q^2)$, evaluated in the $Q^2$-range between $0.5 \div 1$ and $50 ~ (GeV/c)^2$ using our parameterization of $g_1^p$ and $g_2^p$. In Ref.~\cite{SIM02} the leading twist, $\mu_n^{(1)}(Q^2)$, is treated both at next-to-leading ($NLO$) order and beyond any fixed order by adopting available soft gluon resummation ($SGR$) techniques. As for the power corrections, a phenomenological ans\"atz is considered, viz.
 \be
    M_n^{(1)}(Q^2) = \mu_n^{(1)}(Q^2) + a_n^{(4)} {\mu^2 \over Q^2} 
    \left[ {\alpha_s(Q^2) \over \alpha_s(\mu^2)} \right]^{\gamma_n^{(4)}} 
    + a_n^{(6)} {\mu^4 \over Q^4} \left[ {\alpha_s(Q^2) \over 
    \alpha_s(\mu^2)} \right]^{\gamma_n^{(6)}} ~~~~ 
    \label{eq:M1n}
 \ee
where the logarithmic $pQCD$ evolution of the twist-4 (twist-6) contribution is accounted for by an effective anomalous dimension $\gamma_n^{(4)}$ ($\gamma_n^{(6)}$) and the parameter $a_n^{(4)}$ ($a_n^{(6)}$) represents the overall strength of the twist-4 (twist-6) term at the renormalization scale $\mu^2$, chosen to be equal to $\mu^2 = 1 ~ (GeV/c)^2$. In order to fix the running of the coupling constant $\alpha_s(Q^2)$, the updated $PDG$ value $\alpha_s(M_Z^2) = 0.118$ is adopted.

\indent As for the first moment ($n = 1$), the leading twist term, $\delta\mu_1^{(1)}(Q^2)$, does not receive any correction from $SGR$ and at $NLO$ it reads as 
 \be
    \mu_1^{(1)}(Q^2) = {<e^2> \over 2} \left[ \Delta q^{NS} + a_0(Q^2) 
    \right] \left[ 1 - {\alpha_s(Q^2) \over \pi} \right] 
    \label{eq:mu1_first}
 \ee
The non-singlet moment $\Delta q^{NS}$ is taken fixed at the value $\Delta q^{NS} = 1.095$, deduced from the experimental values of the triplet and octet axial coupling constants, with the latter obtained under the assumption of $SU(3)$-flavor symmetry. The values of the singlet axial charge $a_0(\mu^2)$ and of the four higher-twist quantities $a_1^{(4)}$, $\gamma_1^{(4)}$, $a_1^{(6)}$ and $\gamma_1^{(6)}$ are determined by fitting our pseudo-data, adopting the least-$\chi^2$ procedure in the $Q^2$-range between $0.5$ and $50 ~ (GeV/c)^2$. It turns out that the total contribution of the higher twists is tiny for $Q^2 \gsim 1 ~ (GeV/c)^2$, but it is comparable with the leading twist already at $Q^2 \simeq 0.5 ~ (GeV/c)^2$. This means that for $g_1^p$ the onset of {\em global} duality is expected to occur at $Q^2 \simeq 1 ~ (GeV/c)^2$ (cf. Ref.~\cite{duality} for the case of unpolarized structure functions). In our analysis, where the leading and the higher twists are simultaneously extracted, the singlet axial charge (in the $AB$ scheme) is determined to be $a_0(10 ~ GeV^2) = 0.16 \pm 0.09$, which nicely agrees with many recent estimates appeared in the literature. Our value of $a_0$ is therefore significantly below the naive quark-model expectation (i.e. compatible with the well known "proton spin crisis"), but it does not exclude completely a singlet axial charge as large as $\simeq 0.25$.

\indent In case of higher-order moments ($n \geq 3$) both the $NLO$ approximation and the $SGR$ approach have been considered for the leading twist. The comparison of the  corresponding twist analyses shows \cite{SIM02} that, except for the third moment, the contribution of the twist-2 is enhanced by soft gluon effects, while the total higher-twist term decreases significantly after the resummation of soft gluons. Thus, as already observed \cite{SIM00} in the unpolarized case, also in the polarized one it is mandatory to go beyond the $NLO$ approximation and to include soft gluon effects in order to achieve a safer extraction of higher twists at large $x$, particularly for $Q^2 \sim$ few $(GeV/c)^2$.

\indent Finally, the twist decomposition of the polarized Nachtmann moments has been compared with the corresponding one of the unpolarized (transverse) Nachtmann moments obtained in Ref.~\cite{SIM00} adopting the same $SGR$ technique. It turns out~\cite{SIM02} that the extracted higher-twist contribution appears to be a larger fraction of the leading twist in case of the polarized moments. This findings suggests that spin-dependent multiparton correlations may have more impact than spin-independent ones.

\end{document}